\documentclass[useAMS]{mn2e}
\usepackage{graphicx}
\usepackage{slashbox}
\usepackage{multirow} 
\usepackage{longtable}

\title[GHASP: The baryonic Tully-Fisher relation for field galaxies]{GHASP: an H$\alpha$ kinematic survey of spiral and irregular galaxies -- IX. The NIR, stellar and baryonic Tully-Fisher relations.\thanks{GHASP Fabry-Perot data are available at http://fabryperot.oamp.fr}}
\author[S. Torres--Flores et al.]{S. Torres-Flores
$^{1,2,3}$\thanks{Current address. E-mail: storres@dfuls.cl}, B. Epinat$^{1,4}$, P. Amram$^{1}$, H. Plana$^{1,5}$, C. Mendes de Oliveira$^{2}$ \\
$^{1}$Laboratoire d'Astrophysique de Marseille, Universit\'e de Provence \& CNRS,\\
38 rue F. Joliot--Curie, 13388 Marseille, Cedex 13, France\\
$^{2}$Departamento de Astronomia, Instituto de Astronomia, Geof\'isica e Ci\^encias Atmosf\'ericas da USP,\\
Rua do Mat\~ao 1226, Cidade Universit\'aria, 05508-090, S\~ao Paulo, Brazil\\
$^{3}$Departamento de F\'isica, Universidad de La Serena, Av. Cisternas 1200 Norte, La Serena, Chile \\
$^{4}$Institut de Recherche en Astrophysique et Plan\'etologie, Universit\'e de Toulouse \& CNRS,\\
14 avenue Edouard Belin, 31400 Toulouse, France\\
$^{5}$Laborat\'orio de Astrof\'isica Te\'orica e Observacional, Universidade Estadual de Santa Cruz, Ilh\'eus, Brazil}

\begin{document}

\date{}

\pagerange{\pageref{firstpage}--\pageref{lastpage}} \pubyear{2011}

\maketitle

\label{firstpage}

\begin{abstract}

We studied, for the first time, the near infrared, stellar and baryonic Tully-Fisher relations for a sample of field galaxies taken from an homogeneous Fabry-Perot sample of galaxies (the GHASP survey). The main advantage of GHASP over other samples is that maximum rotational velocities were estimated from 2D velocity fields, avoiding assumptions about the inclination and position angle of the galaxies. By combining these data with 2MASS photometry, optical colors, H{\sevensize I} masses and different mass-to-light ratio estimators, we found a slope of 4.48$\pm$0.38 and 3.64$\pm$0.28 for the stellar and baryonic Tully-Fisher relation, respectively. We found that these values do not change significantly when different mass-to-light ratios recipes were used. We also point out, for the first time, that rising rotation curves as well as asymmetric rotation curves show a larger dispersion in the Tully-Fisher relation than flat ones or than symmetric ones. Using the baryonic mass and the optical radius of galaxies, we found that the surface baryonic mass density is almost constant for all the galaxies of this sample. In this study we also emphasize the presence of a break in the NIR Tully-Fisher relation at M$_{H,K}\sim$--20 and we confirm that late-type galaxies present higher total-to-baryonic mass ratios than early-type spirals, suggesting that supernova feedback is actually an important issue in late-type spirals. Due to the well defined sample selection criteria and the homogeneity of the data analysis, the Tully-Fisher relation for GHASP galaxies can be used as a reference for the study of this relation in other environments and at higher redshifts.

\end{abstract}


\section{Introduction}

The Tully-Fisher relation (Tully \& Fisher, 1977) is a direct indication
of a close relationship between the detected baryons and the total
mass in spiral galaxies. The detected baryons consist of the stellar and
gaseous content, i.e. the visible mass, and this sets the luminosity profile
of the galaxy while
the total gravitational mass, which includes the dark matter content (and possibly a component of baryonic dark matter),
sets its rotation velocity. Numerous studies have been carried out to
investigate this relation, crucial in determining
extragalactic distances (e. g. Pierce \& Tully 1988, Tully \& Pierce
2000), in the study of evolution of galaxies (e.g. Puech et al. 2008)
and also in giving constraints on cosmological galaxy formation models
(e. g. Portinari \& Sommer-Larsen 2007).

The Tully-Fisher relation is undoubtedly a crucial test for galaxy evolution models and although it has been the focus of a number of studies, its origin is still being debated. A few authors argue a cosmological origin
(e. g. Avila-Reese, Firmani \& Hern\'andez 1998) while others suggest
that this relation is regulated by star forming processes (e. g. Silk
1997). On the other hand, the Tully-Fisher relation is related to the
stellar populations of galaxies as it is suggested by its steeper slope
when the luminosity is measured in the near infrared (NIR) bands, when compared
to the slope measured in the optical (e.g. Tully \& Pierce, 2000). The use of NIR bands
in the Tully-Fisher relation has shown to be extremely useful, because
NIR bands present lower internal extinction than optical bands (Verheijen
2001) and the mass-to-light ratio is less contaminated by younger stellar
populations, giving a better reflection of the stellar mass of the galaxies.

Luminosities can be converted into stellar masses by
scaling them by a given mass-to-light ratio. Thus, instead of linking rotation
velocities to luminosities, a few authors have chosen to show the correlation of rotation
velocities to stellar masses.  This relation is called stellar
Tully-Fisher relation (e.g. Bell \& de Jong 2001). In order to estimate the whole
baryonic mass, the gaseous component should be added to the stellar
mass. The relation between luminosities and baryonic masses is called the baryonic Tully-Fisher relation. 
This relation has been studied by several authors (e. g. McGaugh 2000, Verheijen
2001, Geha et al. 2006, Bell \& de Jong 2001). The slope of the baryonic TF relation is an important test for galaxy formation models. A steeper slope indicates that the baryonic mass of massive galaxies tends to approximately match the total mass of the galaxy.

To date, most of the works devoted to study the Tully-Fisher relation have used compilations of several galaxy surveys, observed in different ways (H{\sevensize I} line profiles and rotation curves) adding factors of uncertainties due to sample non-homogeneity. Beside the problems of using different samples, the use of certain observational techniques may add other uncertainties in the study of the Tully-Fisher relation. For example, an over or under prediction of the position angle in long-slit observation could produce an erroneous estimate of the maximum velocity of a galaxy, which will be reflected in the Tully-Fisher relation.

In order to avoid the problems listed in the previous paragraph, we have made use of the homogeneous galaxy survey Gassendi HAlpha survey of SPirals (GHASP) to study the NIR, stellar and baryonic Tully-Fisher relations and their implications for the total mass of galaxies. The GHASP survey represents a large effort to constitute a sample of field galaxies in an homogeneous way. First, a strict isolation criterion has been used to insure the isolation of the galaxies. Rather close galaxies have been chosen in order to guarantee a high spatial resolution, compared to H{\sevensize I} surveys. High and low inclination objects have been excluded in order  to minimize uncertainties in the de-projected rotation curve. Second, all GHASP galaxies have been observed using the same instrument; a scanning Fabry-Perot attached to a focal reducer at the 193 cm at Observatoire de Haute Provence (OHP). In addition to obtaining data for the whole sample with the same instrument, which is a great advantage to insure homogeneity, the scanning Fabry-Perot is certainly the most adapted instrument to obtain rotation curves in the most proper way. Because it gives a 2D velocity field, we can obtain the rotation curve (and the maximum velocity) without  any previous assumptions about  the position angle or the inclination like it is the case for long-slit observations. This technique avoids, in this way, a great factor of uncertainty that is common to not be taken into account using other techniques. Third, the data reduction has been performed in an homogeneous way, in order to derive the rotation curves from the velocity fields in the cleanest and most rigorous possible way (see Epinat et al. 2008a,b for details). These three points allowed eliminating (or at least greatly minimizing) several problems that previous studies have encountered. Using rotation curves to obtain the maximum rotation velocity is a more precise way compared to the H{\sevensize I} line width profile technique, used by several others studies. The higher spatial resolution of optical velocity maps, compared to H{\sevensize I}, avoids the problem of missing the maximum velocity because of lack of resolution (beam smearing).

Together with the kinematic information, we have used H and K-band photometry from 2MASS survey, mass-to-light ratios derived from stellar population models, H{\sevensize I} fluxes and H$_{2}$ masses from the literature to perform, for the first time, the NIR, stellar and baryonic Tully-Fisher relations for the GHASP sample.

In \S 2 we describe the data, including the method used to compute the
stellar, gaseous and baryonic masses. In \S 3, we present the results. In
\S 4 we discuss and compare our results with previous works. Finally,
we summarize our main findings in \S 5.

\section{Data}

\subsection{Rotational velocities}

GHASP is the largest sample of spiral and irregular galaxies observed to date using Fabry-Perot techniques. It consists of 3D H$\alpha$ data cubes for 203 galaxies, covering a large range in morphological types and absolute magnitudes. All the GHASP galaxies have been recently reanalyzed in a homogeneous way in Epinat et al. (2008a,b). These authors published velocity fields, monochromatic H$\alpha$ images, dispersion velocity maps, rotation curves and maximum rotation velocities (V$_{max}$) for each galaxy. 

A sub-sample of 93 galaxies has been selected by removing from the sample: 1) galaxies with radial systemic velocities lower than 3000 km s$^{-1}$ (to avoid the effect of the Local Group infall) for which no other individual measurements of distances were available (the references are indicated in Epinat et al. 2008b) and 2) galaxies with inclinations lower than 25 degrees for which the uncertainties on the rotational velocity is comparatively high.

Rotation curves shown in Epinat et al. (2008a,b) present a large variety of shapes (from falling to rising) and degrees of asymmetry. In order to study the influence of the shape of the rotation curves in the Tully-Fisher relation, we have made a classification of our sample in three subsamples, i. e. rising, flat and decreasing rotation curves, which will be described with the letters ``R'', ``F'' and ``D'', respectively.  An ``+'' or `--'' sign has been added if the rotation curve is respectively more or less extended than the optical radius of the galaxy (R$_{25}$). No symbol is added if the radius of the observed rotation curve is barely equal to R$_{25}$.

A decreasing or flat rotation curve displays a clear V$_{max}$. This is not the case for a rotation curve which rises up to the very last observed radius, for which $V_{max}$ is possibly underestimated. This is even worse if the rising rotation curve does not reach the optical radius. In this case, V$_{max}$ was  computed at R$_{25}$ by extrapolating an \textit{arctan} function ($V(r)=V_{0}\times(2/\rm{\pi})\times \rm{arctan}(\it{2r/r_{t}})$, where $r_t$ is the core radius) to the rotation curves. If observed rotational velocities at R$<$R$_{25}$ are higher than the modeled value, the observed values have been used. This could be the case if large scale bumps in the inner parts of rotation curves are present. Table \ref{appendix1} (column 8) presents the results of the rotation curve quality assessments.

The main assumption necessary to derive a rotation curve from the observed velocity field is that rotation motions are dominant and non circular motions are not part of a large-scale pattern. Thus, by construction, a rotation curve provides a measurement, for each radius, of the axi-symmetric component of the gravitational potential well of the galaxy. By consequence, if the motions in the galaxy disk are purely circular, the receding and the approaching sides of a rotation curve should match and the residual velocity field should not display any structure. Once the parameters of the rotation curves are properly computed by minimizing the velocity dispersion in the residual velocity field (Epinat et al. 2008), the remaining residuals are the signature of non-circular motions in or out the plane of the disk (e. g. bars, oval distortions, spiral arms, local inflows and outflows, warps), including the intrinsic turbulences of the gas. To quantify the effects of these non-circular motions on the Tully-Fisher relations we have computed, for each galaxy, two indicators. The first one is based on the asymmetries between both sides of the rotation curves, it quantifies the mean difference of amplitude between the receding and approaching sides of the rotation curve. For each ring centered on the galaxy center, the weighted (absolute) velocity difference between both sides is computed. The weight is provided by the number of bins in each ring. Each bin is an independent velocity measurement on the velocity field, it may be constituted by $\sim$50 pixels for low signal-to-noise region of the galaxy. Depending of the spatial resolution, each ring contains from two to several hundreds bins. Due to the fact that their radius are smaller, the central rings contains a number of pixels lower than the outer ones, their weights is thus smaller. The second indicator is based on the mean velocity dispersion extracted from the residual velocity field. This parameter is quantified by computing the average local velocity dispersion on the whole residual velocity field. To not overestimate the weight of non circular motion in slowly rotating systems with respect to high rotators, both indicators have been normalized by the maximum rotation velocity. We found that both indicators show the same trend on the Tully-Fisher relation, thus we will only illustrate the results using the indicator related to the asymmetries in the rotation curve.

\subsection{Photometry}

We computed the near-infrared magnitudes using 2MASS data (Skrutskie
et al. 2006). 2MASS H and K-band data were available for 83 galaxies of
the GHASP sub-sample defined in section \S2.1. Absolute magnitudes were
obtained using:

\begin{equation}
M_{\rm{H,K}}=m_{\rm{H,K}}+(k_{\rm{H,K}}-A_{\rm{H,K}})-5\times\rm{log}(\textit{D})-25
\label{dis}
\end{equation}

Distances (\textit{D}) were taken from Epinat et al. (2008b). They are computed from the systemic velocities (from the NED database) corrected from Virgo infall and assuming H$_{0}$=75 km s$^{-1}$ Mpc$^{-1}$, except when accurate distance measurements where available (references are listed in Epinat et al. 2008b). The magnitudes m$_{\rm{H,K}}$ have been computed using the flux within the isophote of 20 mag arcsec$^{-1}$ (where uncertainties were taken from 2MASS). We corrected the magnitude for Galactic extinction using the Schlegel maps (Schlegel et
al. 1998). k-corrections (k$_{\rm{H,K}}$), extinction due to the inclinations (A$_{\rm{H,K}}$) and seeing
were applied using the method given in Masters et al. (2003). Columns 1,
2 and 3 in Table \ref{appendix1} list the name, H and K-band absolute
magnitudes for the sample. K-band luminosities were estimated using
L$_{K}$=10$^{-0.4(M_{K}-3.41)}$, where the K-band absolute Solar magnitude of 3.41 was taken from Allen (1973).

Given the homogeneity of the \textit{Sloan Digital Sky Survey} (\textit{SDSS}), g and r-band optical magnitudes were extracted from this database. Moreover, most of the mass-to-light ratio recipes use B-V colors as an input parameter. For this reason, we have converted g and r-band data into B and V-band magnitudes by using the recipes given in Lupton (2000). We have extracted the optical size for each galaxy from the \textit{SDSS}. In this case, we used the parameter isoA (in the r-band), which corresponds to the diameter of the isophote where the disk surface brightness profile drops to 25 mag arcsec$^{-2}$. These values were converted in radii (in kpc) by using the distance to each galaxy.

In order to compute the mass-to-light ratio for GHASP galaxies, g-r colors were corrected by Galactic extinction using the values given in the \textit{SDSS} database and then converted into B-V colors. \textit{SDSS} colors were available for 45 galaxies from which we removed five objects for which their magnitudes and optical radius are obviously incorrect (compared with the optical radius, and magnitudes, given in HyperLeda). For other six galaxies, there were no radius measurement.  All the analysis including the radius of the galaxies were thus performed with 34 objects.

B-band luminosities were estimated by using L$_{B}$=10$^{-0.4(M_{B}-5.48)}$, where the B-band absolute Solar magnitude of 5.48 was taken from Binney \& Merrifield (1998).

\subsection{Mass-to-light ratios and stellar masses}

The main uncertainty in the study of the stellar and baryonic
Tully-Fisher relations states in the stellar mass-to-light disk ratio
$\Upsilon_{\star}$. Two main methods to estimate this parameter are used. Spano et al. (2008) have modeled the stellar mass
distribution of rotation curves, by scaling the R-band surface brightness profile to the rotation curve, obtaining an estimation of $\Upsilon_{\star}$. Bell \& de Jong (2001) have used stellar populations
synthesis models to predict a relation between the colors of galaxies
and their $\Upsilon_{\star}$. Although both approaches attempt to compute the same
physical parameter, several authors have shown that surprisingly there seems to be 
no clear correlation between the $\Upsilon_{\star}$ obtained from these two methods
(e.g. Barnes et al. 2004). Other authors have invoked the modified
Newton dynamics (MOND) to obtain the $\Upsilon_{\star}$ and study its
implication on the baryonic Tully-Fisher relation (McGaugh 2005). In
this work, we have estimated $\Upsilon_{\star}$ and stellar masses using
stellar population synthesis models recipes. One of them consists of simply
fixing the value of $\Upsilon_{\star}$. We have compared our results  with
other works available in the literature following the stellar populations
synthesis models given by Bell \& de Jong (2001, B\&J), Bell et al. (2003,
BE) and Portinari et al. (2004, PO) (equations \ref{eq1} and \ref{eq1b},
\ref{eq2} and \ref{eq2b} and \ref{eq3} respectively).

\begin{equation}
M_{\star}^{\rm{B\&J}}=10^{-0.692+0.652(B-V)} L_{K} 
\label{eq1}
\end{equation}

\begin{equation}
M_{\star}^{\rm{B\&J}}=10^{-0.994+1.804(B-V)} L_{B}
\label{eq1b}
\end{equation}

\begin{equation}
M_{\star}^{\rm{BE}}=10^{-0.206+0.135(B-V)} L_{K} 
\label{eq2}
\end{equation}

\begin{equation}
M_{\star}^{\rm{BE}}=10^{-0.942+1.737(B-V)} L_{B}
\label{eq2b}
\end{equation}

\begin{equation}
M_{\star}^{\rm{PO}}=10^{0.730[(B-V)-0.600]-0.115} L_{K} 
\label{eq3}
\end{equation}

B\&J and BE suggested an uncertainty of 0.1 dex in the $\Upsilon_{\star}$ estimation. The adopted uncertainty in $\Upsilon_{\star}$ is larger than the uncertainties of the optical colors. We have adopted the same uncertainty for the PO relation. B\&J used a scaled Salpeter initial mass function (IMF). PO used a Salpeter IMF, with masses ranging between 0.1 and 100 M$_{\odot}$. These
models are available for several colors (to estimate $\Upsilon_{\star}$) and also for the luminosity in several bands (to estimate the stellar mass). We have converted \textit{SDSS} g-r colors into B-V colors to obtain the $\Upsilon_{\star}$ parameter by using the recipes listed above. Stellar masses were calculated using the K-band luminosities. This band is more likely to be reflective of the stellar mass of galaxies. In order to compare stellar masses derived from the K-band luminosities, we have also used the B-band luminosity, despite the fact that this photometric band could be contaminated with the emission of
young stars. We removed from this analysis galaxies for which no \textit{SDSS}
colors were available in the literature. Therefore, in the stellar and
baryonic analysis, we were left with 45 galaxies in total. K-band luminosities were estimated
using L$_{K}$=10$^{-0.4(M_{K}-3.41)}$ (see section 2.2).

Stellar masses were also estimated using a fixed mass-to-light ratio
following McGaugh et al. (2000, \textit{MG}). These authors defined
the mass-to-light ratio in the K-band as $\Upsilon_{\star}=0.8
M_{\odot}/L_{\odot}$. In this case, the stellar mass is simply:

\begin{equation}
M_{\star}^{MG}=\Upsilon_{\star}L_{K}
\label{eq3b} 
\end{equation}

It is interesting to note that Gurovich et al. (2010) estimated the
$\Upsilon_{\star}$ for a sample of local galaxies by modeling their
stellar population histories. These authors did not find differences
in the Tully-Fisher relation when the stellar masses were computed by using
the modeled $\Upsilon_{\star}$ or when this parameter was fixed to
$\Upsilon_{\star}$=0.8 (McGaugh et al. 2000).

\subsection{Baryonic masses}

The mass of a galaxy is constituted of its content in stars and stellar remnants, gas (neutral,
molecular and a negligible component of ionized gas), dust (usually negligible)
and dark matter. The baryonic mass is the sum of the stellar and gas contents.
The total mass in gas, M$_{gas}$, is:

\begin{equation}
M_{gas}=M_{H\sevensize I}+M_{He}+M_{H_{2}}+(M_{HII})
\end{equation}

where M$_{H\sevensize I}$ is the neutral gas, M$_{He}$ is the mass in
helium and metals, M$_{H_{2}}$ is the mass in molecular hydrogen and
M$_{HII}$ is the (negligible) mass in ionized hydrogen.

In order to obtain the baryonic mass for the GHASP sample we have
calculated the observed HI mass for each galaxy using the corrected
21-cm line flux taken from HyperLeda (Paturel et al. 2003). Fluxes were converted into mass
using the relation:

\begin{equation}
M_{H{\sevensize I}}=2.356\times10^{5}F_{HI} D^2 
\label{eqhi}
\end{equation}

where D is the distance to the galaxy in Mpc, and F$_{H{\sevensize
I}}$ is the H{\sevensize I} flux in Jy km$^{-1}$. 

To take into account the correction for helium and metals in the gas
content (e. g. McGaugh et al. 2000), M$_{He}$ is related to the HI
mass through:

\begin{equation}
M_{He}=0.4M_{H \sevensize I}
\end{equation}

The H$_{2}$ mass has been computed following the formula given by McGaugh \& de Blok (1997), using the morphological type of the galaxies
(Young \& Knesek, 1989).

\begin{equation}
M_{H_{2}}=M_{H \sevensize I}(3.7-0.8T+0.043T^{2})
\end{equation}

Nevertheless, to avoid uncertainties linked to our bad knowledge in
the H$_{2}$ content, the baryonic mass studied in this paper does not
include H$_{2}$, except when it is explicitly mentioned.  The baryonic
mass, M$_{bar}$, is defined as:

\begin{equation}
M_{bar}=M_\star+M_{gas} 
\end{equation}

where M$_\star$ is the total stellar mass. Uncertainties in the baryonic mass are the results of the quadratic sum of the uncertainties in stellar masses and uncertainties in the H {\sevensize I} masses, which were taken from HyperLeda.

We have compared the baryonic mass to the total dynamical mass for each galaxy of our sample. Although almost the whole baryonic mass is approximately within the optical radius, the total dark matter content is not, thus we compute the dynamical total mass only within the optical radius. To estimate the total dynamical mass, we assumed the mass has a spherical distribution, which is likely the case for the dark halo component, as supported by observations (e.g. Ibata et al. 2001) and N-body simulations (e.g. Kazantzidis et al. 2010) by using:

\begin{equation}
M(R)=\alpha R \times V_{max}^{2}/G
\end{equation}

where $\alpha$ is a parameter depending on the mass profile distribution (equal to one for an uniform distribution). To compute M(R) we have used the r-band optical radius from \textit{SDSS}, as tabulated in the Appendix. In order to obtain an estimation of the dark matter content at the optical radius, we have subtracted the baryonic mass (as estimated in \S 2.5) from the dynamical total mass within the optical radius.

\subsection{Fitting method}

Galaxies having the same rotational velocity do not necessarily have the same luminosity (or reciprocally), thus the observed Tully-Fisher relation presents a dispersion which may be produced by intrinsic properties of galaxies. Beside this dispersion, uncertainties in magnitudes/masses and rotational velocities should be taken into account when the slope and zeropoint of this relation are computed (see Hogg et al. 2010 for details about fitting straight lines). Several efforts have be performed to fit straight lines to fundamental relations, taken into account together the uncertainties in both axis and the intrinsic dispersion of the relation (e. g. Tremaine et al. 2002, Weiner et al. 2006). In this paper, we have followed the prescription given by Tremaine et al. (2002), by adding (in quadrature) a dispersion factor to the uncertainties estimation of the NIR magnitudes, stellar and baryonic masses. The value of the dispersion factor is chosen in order to reach a $\chi^{2}$ of unity per degree of freedom. To fit the Tully-Fisher relation, we used linear relation of the form:

\begin{equation}
y=\alpha x+\beta
\end{equation} 

where, $y=M_{H,K}$ and $y=log(M/M_{\odot}$), for the NIR and stellar/baryonic Tully-Fisher relation, respectively, and $x=log(V_{max}/km s^{-1})$. To obtain the slope and the zerpoint of this relation, we used the task FITEXY (Press et al. 1992).

\begin{figure*}
\includegraphics[scale=0.55]{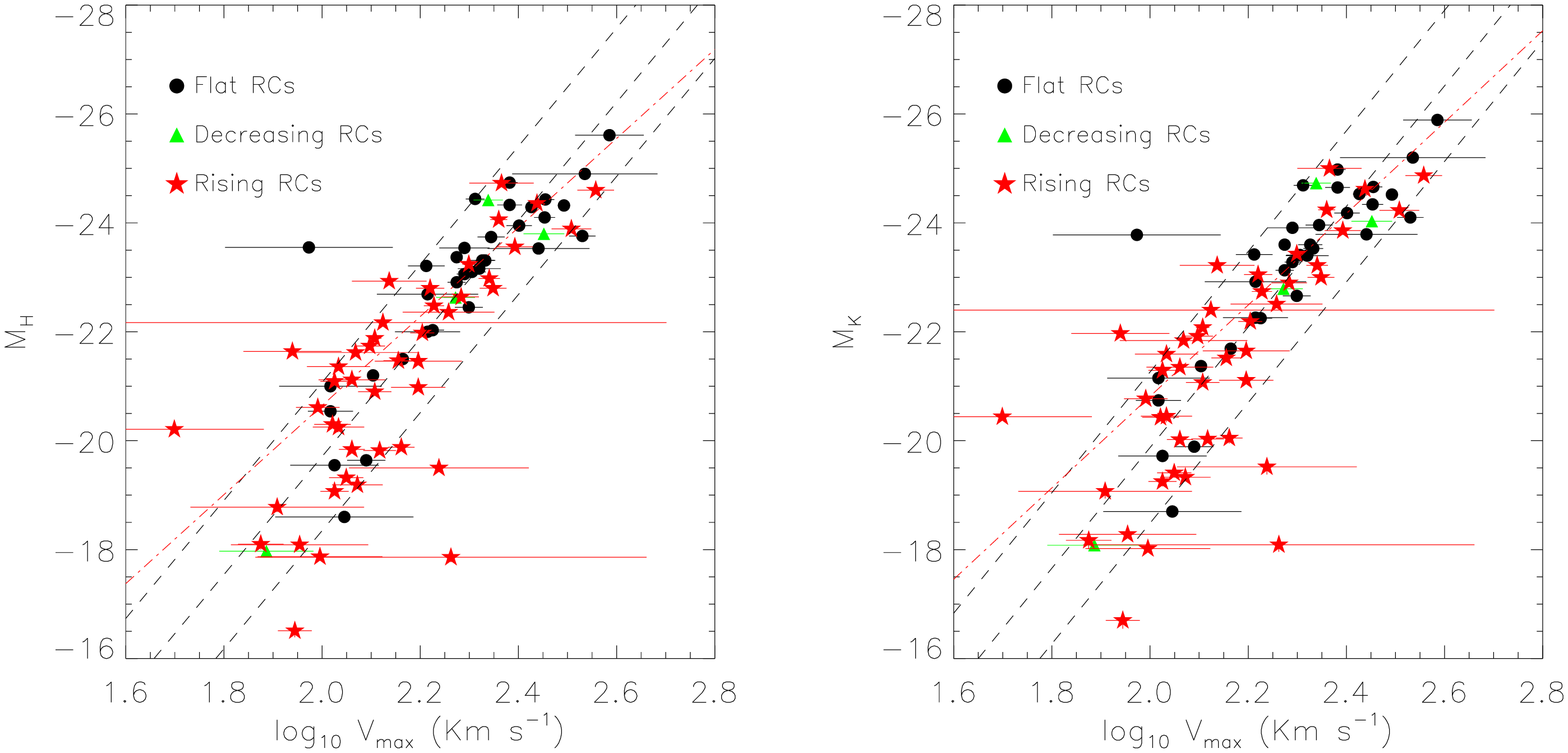}
\includegraphics[scale=0.55]{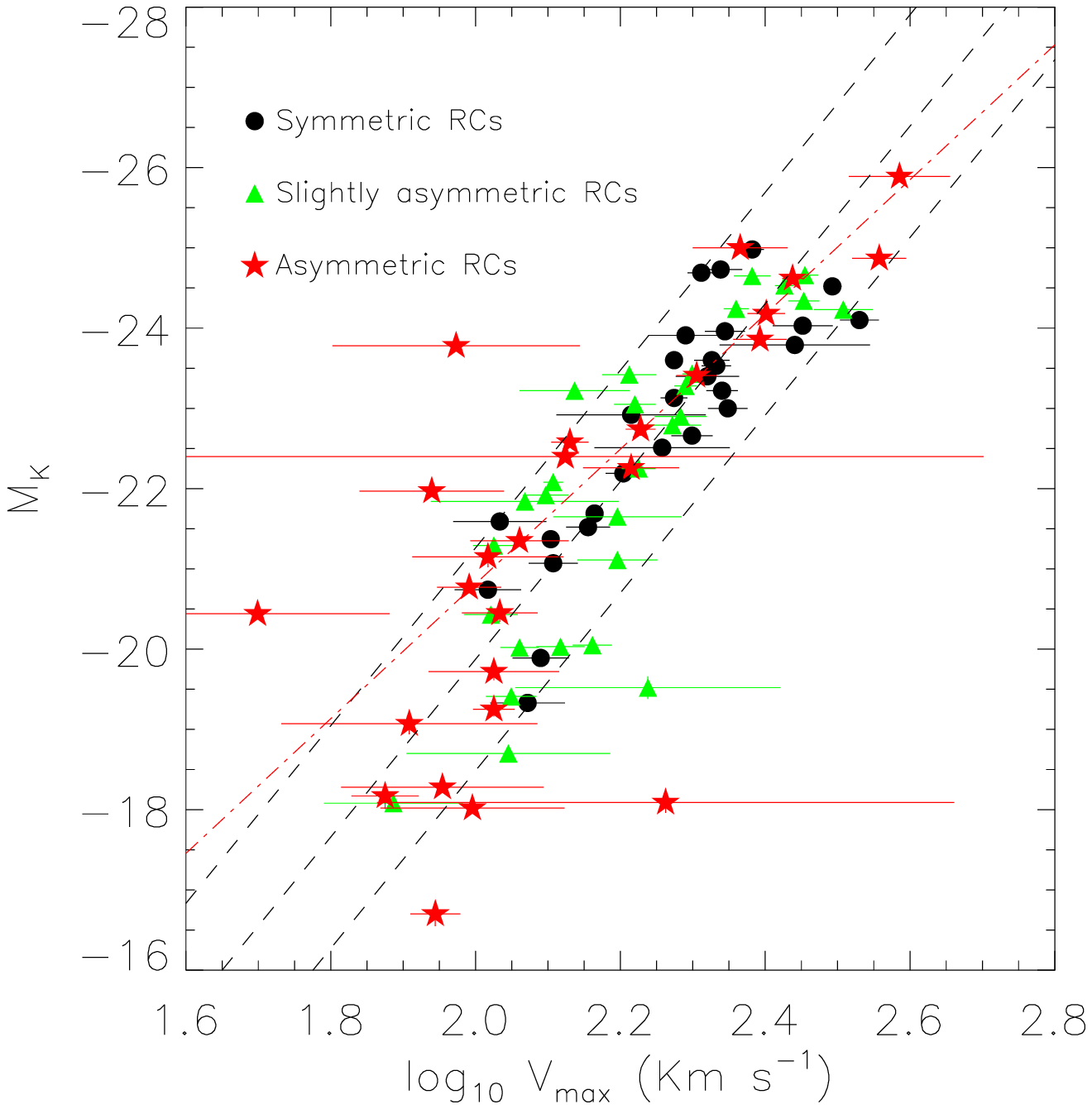}
\caption{Upper left and right panels: H-band and K-band Tully--Fisher relation for the GHASP sample. The dashed line represents the best fit on all the data (with one-$\sigma$ in the zeropoint), following $M_{H,K}=(\beta\pm\sigma_{\beta})+(\alpha\pm\sigma_{\alpha})logV_{max}$ (see Table \ref{table2}). The red dashed-dotted line represents the fit for galaxies with M$_{K}<$-20 (see \S3.1). Black dots, green triangles and red stars represent flat,  decreasing and rising rotation curves. Bottom panel: K-band Tully-Fisher relation on which galaxies have been distinguished by their ratio between non-circular and circular motions using the asymmetry between both sides of the rotation curve. The sample has been divided in three classes: class I (black dots) is attributed to the galaxies showing the lower non-circular motions; class III (red stars) to the galaxies exhibiting the higher non-circular motions and class II (green triangle) to mild non-circular motions.}
\label{BTF_GHASP_fig2a}
\end{figure*}

\section{Results}

In Table \ref{appendix1} we list the NIR magnitudes and the different
rotational velocities that we obtained, from the observations and from
the arctan model (see \S2.1). Columns 1, 2 and 3 list the name,
H and K-band magnitudes. Column 4 shows the observed maximum rotational velocity
for each galaxy. Column 5 gives the modeled velocity at R$_{25}$. Column 6
corresponds to the rotational velocity used in the TF relation. Finally,
in column 7 we classify the shape of rotation curves of the GHASP sample as shown in \S 2.1.

In Table \ref{appendix2} we list the different determinations of mass-to-light ratios ($\Upsilon_{\star}$) and masses for each galaxy. Columns 1, 2 and 3 indicate respectively the name, the radius of the galaxies (taken from \textit{SDSS}) and the B-V colors (transformed from g-r colors). In columns 4, 5 and 6 we list the mass-to-light ratios calculated from equations \ref{eq1}, \ref{eq2} and \ref{eq3}, respectively. Values for the stellar masses are shown in columns 7, 8 and 9, following B\&J, BE and PO, respectively. In column 10, we list the M$_{H{\sevensize I}}$+M$_{He}$, where M$_{H{\sevensize I}}$ is calculated using the observed H{\sevensize I} mass for the GHASP sample. Column 11 corresponds to the baryonic masses excluding the H$_{2}$ content (M$_{\star}$+M$_{H{\sevensize I}}$+M$_{He}$), M$_{\star}$ is here calculated following BE given in column 8. 

\subsection{H and K-band TF relations}

In Fig. \ref{BTF_GHASP_fig2a} (upper panels) we plot the Tully-Fisher relations for the H and K-band (left and right panels, respectively). In both panels, flat, decreasing and rising rotation curves are indicated by black dots, green triangles and red stars, respectively. Galaxies having a rising rotation curve show a large dispersion on the Tully-Fisher relation, while most of the flat rotation curves lie on the relation. This may simply reflect the scatter in the determination of V$_{max}$ for rising rotation curves, for which V$_{max}$ may be uncertain. Alternatively, this might indicated that rising rotation curves, that are usually dark matter dominated galaxies, show an intrinsic scatter in the Tully-Fisher relation. In the bottom panel of Fig. \ref{BTF_GHASP_fig2a} we plot the K-band Tully-Fisher relation in which we divided the sample by their asymmetries in the rotation curve. Galaxies displaying the largest non-circular motions (red stars) lie preferentially in the low velocity/luminosity region of the plot and present a larger scatter than galaxies less affected by non circular motion (black dots). The conclusion is that non-circular motions contribute to the scatter in the NIR Tully-Fisher relation, at least in the low luminosity (and mass) region of the plot.

Inspecting Fig. \ref{BTF_GHASP_fig2a}, we observe a break
in the Tully-Fisher relation at M$_{H,K}\sim$-20. This effect has already been
noted by McGaugh et al. (2000), Gurovich et al. (2004) and Amor\'in et
al. (2009). In order to quantify this break, we have applied a fit (as discusses in \S 2.5) to all galaxies (black dashed line) and to the
galaxies with M$_{K}<$-20 (red dotted-dashed lines). For the first case,
we have derived the followings equations:

\begin{equation}
M_{H}=(1.97\pm1.36)-(10.84\pm0.61)[log(V_{max})] 
\label{mh}
\end{equation}

where we use a dispersion factor of 0.75 in the H-band magnitude.

\begin{equation}
M_{K}=(2.27\pm1.39)-(11.07\pm0.63)[log(V_{max})]
\label{mk}
\end{equation}

where we use a dispersion factor of 0.76 in the K-band magnitude.

For galaxies with M$_{K}<$--20, we have derived:

\begin{equation}
M_{H}=(-4.29\pm1.14)-(8.18\pm0.50)[log(V_{max})] 
\label{mh20}
\end{equation}

where we use a dispersion factor of 0.47 in the H-band magnitude.

\begin{equation}
M_{K}=(-4.02\pm1.17)-(8.39\pm0.52)[log(V_{max})]
\label{mk20}
\end{equation}

where we use a dispersion factor of 0.49 in the K-band magnitude.

\begin{figure*}
\includegraphics[scale=0.55]{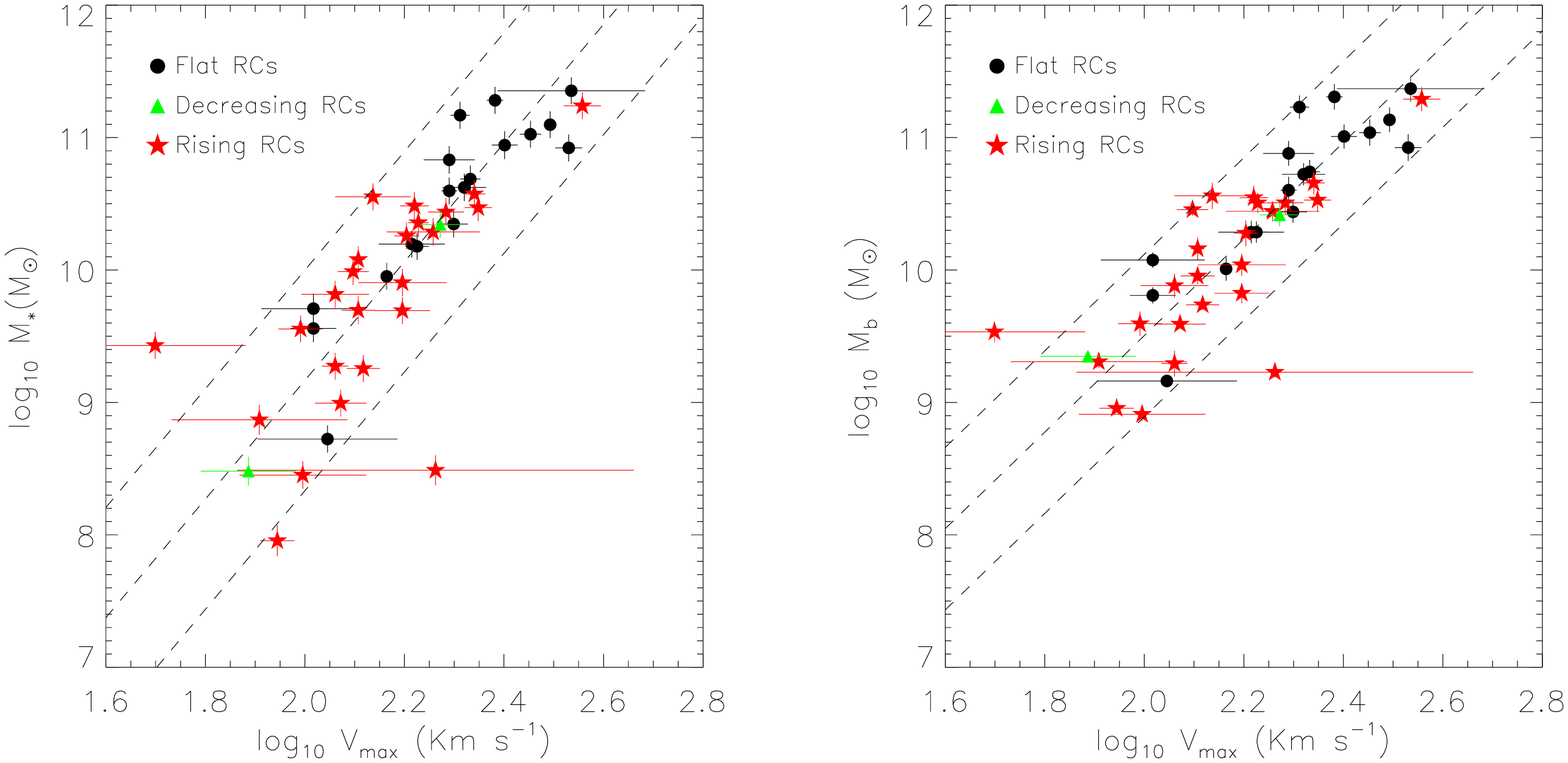}
\caption{Stellar (left panel) and baryonic (right panel) Tully--Fisher relation for the GHASP sample. The dashed lines represent the best fit on the data (see \S 2.5). Flat, decreasing and rising rotation curves are represented by circles, triangles and stars, respectively.}
\label{BTF_GHASP_fig2b}
\end{figure*}

In the equations above we have included the one-sigma uncertainties in the slope and zero point. Three galaxies ($\sim$4$\%$ of the sample) are 1$\sigma$ off the Tully-Fisher relation. These galaxies could have slightly
too high magnitudes for their rotational velocities or too low
rotational velocities for their magnitudes (upper left region in
each panel of Fig. \ref{BTF_GHASP_fig2a}). In spite of the rotation
curves of these objects reach the optical radius R$_{25}$, two of
them have rising rotation curves (red stars in the left panels of
Fig. \ref{BTF_GHASP_fig2a}). In this sense, we can not exclude that both
galaxies could have higher rotational velocities than the observed values,
placing both galaxies on the TF relation. On the other hand, one galaxy
has a flat rotation curve (black dot in Fig. \ref{BTF_GHASP_fig2a}),
meaning that this object already reached its maximum rotational
velocity. Two possible scenarios could explain the high near-infrared
magnitude of this galaxy. AGN activity, that could enhance the
near-infrared magnitude of Seyfert 1 galaxies (Riffel et al. 2009)
and/or the contribution of TP-AGB stars into the K-band luminosity
(Maraston 1998). TP-AGB stars are often not taken into account in the
models and they are quite important specially for stellar populations with ages below 1 Gyr. A detailed study of this galaxy is out of the scope of this paper.

Taking into account the dispersions, the slopes of the H and K-band
TF relations are similar, being slightly steeper in the K-band. Table
\ref{table2} summarizes the fit parameters
(denoted by $\alpha$ for slopes and $\beta$ for the zero points) found
for the TF relation in H and K-bands. In the same table we list values
of $\alpha$ and $\beta$ found in the literature.

\subsection{The stellar TF relation}

In order to compare different estimators of the stellar mass available
in the literature, we have calculated the slope and zero points of the
stellar TF relations of our GHASP sample using different prescriptions for
mass determination given by B\&J, BE, PO and  MG (see equations \ref{eq1},
\ref{eq2}, \ref{eq3} and \ref{eq3b}). Table \ref{table3} summarizes the
fit parameters using the different stellar mass estimators. Slopes from 
different estimators are
consistent within 1$\sigma$. In the
following, we used the BE results to study the stellar and baryonic TF
relations (the BE models are updated versions of the B\&J models).

In Fig. \ref{BTF_GHASP_fig2b} (left panel), we show the stellar TF relation for the GHASP sample. The central dashed line follows equation \ref{stellar} (and one-$\sigma$ in the zeropoint) and represents the best fit on the data. In this case, we use a dispersion factor of 0.31 dex in the stellar mass.

\begin{equation}
M_{\star}=10^{(0.21\pm0.83)}V_{max}^{(4.48\pm0.38)}
\label{stellar}
\end{equation}

We found a slope of 4.48$\pm$0.38 (equation \ref{stellar}) when we
calculated the stellar masses following BE. B\&J found
a slope of 4.4$\pm$0.2 for the stellar TF relation, for a sample of
galaxies in the Ursa Major cluster.

As previously found in the H and K-band TF relation, low-mass galaxies
in Fig. \ref{BTF_GHASP_fig2b} (with stellar masses lower than 10$^{9}$
M$_{\odot}$) lie below the relation defined by high-mass spirals. By using cosmological hydrodynamical simulations, de Rossi et al. (2010) suggested that SN feedback is the main responsible for this behavior in low-mass systems.

\begin{figure}
\centering
\includegraphics[width=\columnwidth]{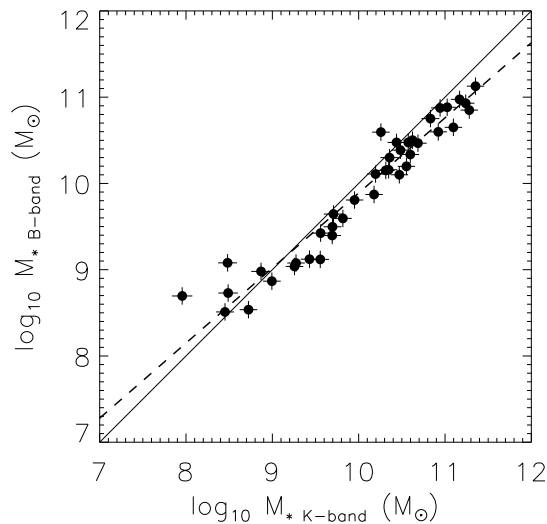}
\caption{Stellar masses derived by using BE models. In the X-axis masses
were calculated using the K-band luminosities while in the Y-axis,
stellar masses were calculated using the B-band luminosities. The dashed
line represents the best fit on the data and the continuous line
represents \textit{y=x}.} \label{BTF_GHASP_fig6} 
\end{figure}

In Fig. \ref{BTF_GHASP_fig6} we plot stellar masses derived from the
K and B-band luminosities, following the recipes of BE. We found that
the B-band luminosities slightly overestimate (underestimate) the stellar mass for the low-mass (high-mass)
galaxies, when it is compared with the value derived from the K-band
luminosity. Such an overestimation (and underestimation) for the stellar masses of low-mass (and high-mass) galaxies
could strongly affect the slope of the baryonic Tully-Fisher relation,
adding a ``band-dependence effect". The same trend was found when we use the recipes given in BJ and PO.

\subsection{The baryonic TF relation}

In Fig. \ref{BTF_GHASP_fig2b} (right panel), we plot the baryonic TF
relation for GHASP. The central dashed line represents the best fit for the data and one-$\sigma$ in the zeropoint (see equation \ref{bary}). In this case, we use a dispersion factor of 0.21 dex in the baryonic mass. The shapes of the rotation curves are represented by different symbols (circles, triangles and stars
indicated flat, decreasing and rising rotation curves). 

\begin{equation}
M_{bar}=10^{(2.21\pm0.61)}V_{max}^{(3.64\pm0.28)}
\label{bary}
\end{equation}

It is interesting to note that low-mass galaxies now lie on the same relation defined by high-mass galaxies. This fact is attributed to the inclusion of the gaseous mass into the stellar budget. In this plot, the stellar mass was estimated from BE. The slope of the baryonic TF that we derive for the GHASP sample is 3.64$\pm$0.28 (which is in agreement with the slope obtained from an unweighted bisector fit, i.e. 3.58$\pm$0.37). As done
for the stellar mass TF relation, we obtained the slope and zero point
of the baryonic TF relation when the stellar mass was calculated using
B\&J, BE, PO and MG. These values are listed in Table \ref{tablehenri},
where $\alpha$ and $\beta$ correspond to the zero points and slopes,
respectively. In Table \ref{tablehenri} we have included the resulting
slopes and zero points obtained when H$_{2}$ masses for the 
galaxies are included in the
baryonic budget. We found that the slope of the baryonic TF relation,
when H$_{2}$ is included, does not change significantly.

Note, in particular, that the use of the Bell \& de Jong (2001) or Bell et al. (2003) mass-to-light recipes, on the GHASP sample, results in a very similar slope of the baryonic TF relation (lines 1 and 9 of Table \ref{tablehenri}).

In Table \ref{tablehenri} we also list the fit parameters for the
baryonic Tully-Fisher relation when the stellar masses were derived
from the B-band luminosities (and the gaseous mass was corrected by the H$_{2}$
gas mass). In this case, the slope is shallower ($\alpha=3.25\pm0.29$) than
in the case when stellar masses are computed from the K-band luminosities.

\begin{figure}
\centering
\includegraphics[width=0.93\columnwidth]{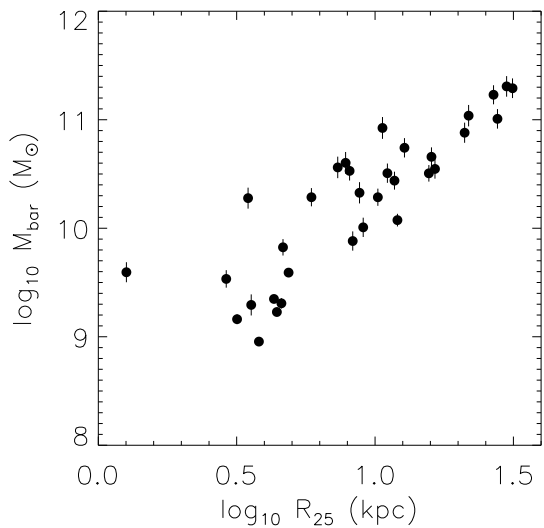}
\caption{Correlation between R$_{25}$ and the baryonic mass of GHASP galaxies.}
\label{r25bary}
\end{figure}

In Fig. \ref{r25bary}, we show that, even with some scatter in the
relation, the baryonic mass $M_{bar}$ grows almost linearly with the optical
galactic radius R$_{25}$ in log units ($M_{bar}=\eta R_{25}^\gamma$). We
used a weighted bisector least square fit to obtain the dependence between
these parameters (we used a bisector fit given that the \textit{SDSS} database does not quote errors in the radius). We found $M_{bar}=(7.88\pm0.01)R_{25}^{(2.39\pm0.01)}$, where $log(\eta)=7.88\pm0.01$. These results suggest that the baryonic
mass density, defined by $\Sigma_{b}$=$M_{bar}/R_{25}^{2}$, depends
weakly on the sizes of the galaxies. More precisely, $\Sigma_{b} \propto R_{25}^{0.4}$.

\begin{table}
\centering
\begin{minipage}[t]{\columnwidth}
\caption{H and K-band Tully-Fisher fit parameters}
\begin{tabular}{lcc}
\hline
  &\multicolumn{2}{c}{H-band}\\
\multicolumn{1}{c}{References}\footnotetext{Comparison between the Tully-Fisher fit parameters (slopes and zero points are represented by $\alpha$ and $\beta$, respectively) for this work and other studies available in the literature.} & $\alpha$&$\beta$\\
 \hline
This work &-10.84${\pm}$0.61&1.97${\pm}$1.36  \\
Masters et al. (2008)   &-9.02 &0.06 \\

\hline
 & \multicolumn{2}{c}{K-band}\\
&$\alpha$&$\beta$\\
\hline

This work  &-11.07${\pm}$0.63&2.27${\pm}$1.39  \\
Masters et al. (2008) &-10.02& 2.37 \\
Courteau et al. (2007)    &-9.29& 1.24  \\
Karachentsev et al. (2002)     &-9.02&...        \\
Verheijen (2001)     &-10.60&3.45  \\
Rothberg et al. (2000)      &-8.78&-1.22  \\

\hline
\vspace{-0.8cm}
\label{table2}
\end{tabular}
\end{minipage}
\end{table}

\begin{table}
 \centering
 \begin{minipage}{\columnwidth}
  \caption{Stellar Tully-Fisher fit parameters for the GHASP sample}
  \begin{tabular}{lcc}
  \hline
   Model\footnotetext{Tully-Fisher fit parameters obtained from the ordinary least square bisector fit. Slopes and zero points are represented by $\alpha$ and log($\beta$), respectively.}&$\alpha$&log($\beta$)\\
 \hline
Bell \& de Jong (2001)     &4.68$\pm$0.40   &-0.27$\pm$0.88   \\
McGaugh et al. (2000)     &4.47$\pm$0.54   &0.19$\pm$1.23    \\
Bell et al. (2003)                &4.48$\pm$0.38   &0.21$\pm$0.83    \\
Portinari et al. (2004)        &4.70$\pm$0.40  &-0.11$\pm$0.89    \\
\hline
\vspace{-0.8cm}
\label{table3}
\end{tabular}
\end{minipage}
\end{table}

\subsection{Total mass versus baryonic mass}

\begin{figure}
\centering
\includegraphics[width=\columnwidth]{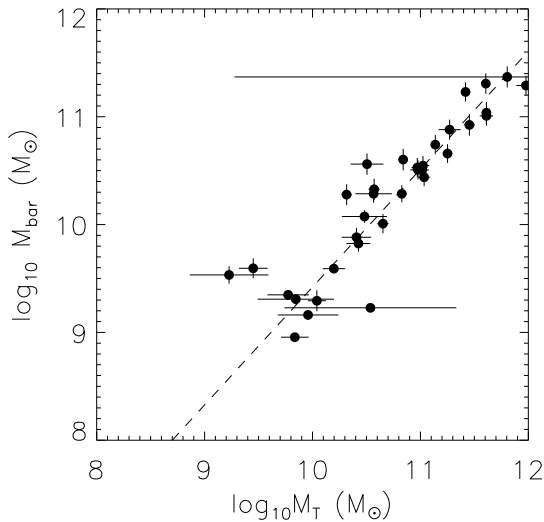}
\caption{Comparison between dynamical mass at R$_{25}$ and baryonic mass. Dashed line represent a bisector least square fit.}
\label{totmassbary}
\end{figure}

In Fig. \ref{totmassbary} we have plotted the baryonic mass
versus the total mass (computed at the optical radius). This plot shows
explicitly the relation between the baryonic and the dark matter
content, given that the masses were computed totally 
independently along the two axis. We fit a least square fit to the data (taken into account errors in both axis), obtaining
$M_{bar}=10^{(-1.51\pm0.42)}M_{T}^{(1.09\pm0.04)}$. Given that the slope of this
relation is close to one, we have estimated, on average, how much mass we
need to add to the baryonic mass to reach the total mass at R$_{25}$. We
have used the ratio (M$_{T}$-M$_{bar}$)/M$_{T}$ at R$_{25}$ (which is
an indicator of the dark matter content inside R$_{25}$) finding a mean
value of 0.66 with a standard deviation of 0.17. In this context, we found
that, on average, the baryonic mass at R$_{25}$ is about $\sim$34$\%$ that 
of the total mass of the galaxies. 

When we divided our sample in low-mass and high-mass galaxies, we found that in the first case (GHASP galaxies having rotational velocities lower than 150 km s$^{-1}$) that 29$\%$ of their total mass corresponds
to baryonic mass. When we consider intermediate and high-mass galaxies (V$_{max}$$\ge$150 km s$^{-1}$), this percentage increase
to 37$\%$ of the total mass. In general, these results are in agreement
with previous works (e.g. Persic et al. 1996).

\begin{figure}
\includegraphics[width=\columnwidth]{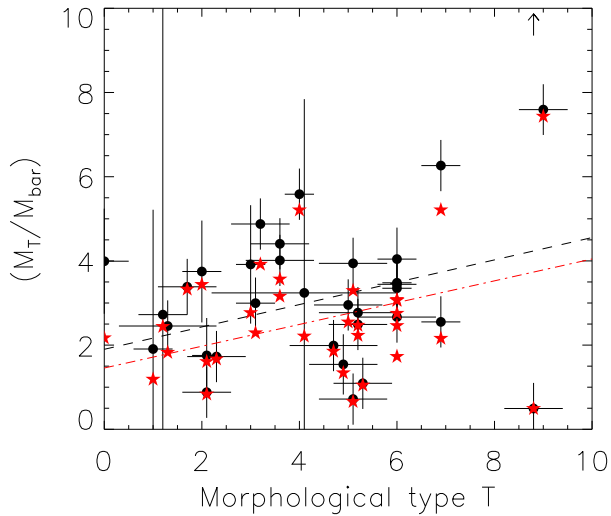}
\caption{Total-to-baryonic mass ratio versus morphological type. Filled dots represent the gaseous mass used in this work. Stars represent the gaseous mass corrected by H$_{2}$. The least square fit of the filled circles is shown by a dashed lines and the fit of the stars by a dashed-dot-dashed line.}
\label{BTF_GHASP_fig5}
\end{figure}

In Fig. \ref{BTF_GHASP_fig5} we plot the total-baryonic mass ratio versus the morphological type. Red stars indicated that molecular hydrogen has been taken into account in the baryonic mass estimation and the red dashed-dotted line is its corresponding fit (as described in \S 2.5 and excluding the outlier with M$_{T}$/M$_{bar}\sim$20). Black dots represent the baryonic mass with no contribution from molecular hydrogen (and the black dashed line represent the fit in this case). From Fig.  \ref{BTF_GHASP_fig5} we can note that the dark halo has a larger contribution to the total mass in late-type spirals (or the contribution of the baryonic mass is less important) than in early-type spirals. When the H$_{2}$ is taken into account (red stars), the trend is the same. We found one galaxy (UGC 6628) for which the M$_{T}$/M$_{bar}$ ratio was $\sim$20 (indicated by an arrow). In this case, the large uncertainty in the maximum rotational velocity (see Table \ref{appendix1}) can affect the determination of the total mass, placing this galaxy outside the relation defined by the others galaxies.

We caution the reader that the baryonic matter fractions derived here are  dependent on the many assumptions made deriving the color-$\Upsilon_{\star}$ prescriptions. Most notably, the zero-points of the B\&J and BE prescriptions were derived using a maximum disk constraint from rotation curves. Therefore, the baryonic fraction derived here are best seen as upper limits, the baryon fractions could be lower if $\Upsilon_{\star}$ prescriptions were normalized lower.

\begin{table*}
\centering
\begin{minipage}[c]{0.7\textwidth} 
 \caption{Fits parameters of the baryonic Tully-Fisher relation}
 \begin{tabular}{ c l l c c c  c c c  }
 \hline

&   &\multicolumn{1}{c}{Authors}  &\multicolumn{2}{c}{Coefficients} & RC & Gas mass \\
\backslashbox{Model} & & &$\alpha$  & $\beta$ &  \\
\hline
(1)& \vline&\multicolumn{1}{c}{(2)}&(3)&(4)&(5)&(6)\\
\hline
\multirow{6}{*}{BJ01}       
  &   \vline & GHASP sample (this work, L$_{K}$)          & 3.72$\pm0.29$&2.02&H$\alpha$& H{\sevensize I}+He\\
  &   \vline & GHASP sample (this work, L$_{K}$, H$_{2}$) & 3.69$\pm0.29$&2.18&H$\alpha$& H{\sevensize I}+He+H$_{2}$\\
  &   \vline & GHASP sample (this work, L$_{B}$, H$_{2}$) & 3.27$\pm0.29$&3.03&H$\alpha$& H{\sevensize I}+He+H$_{2}$\\
  &   \vline & Bell \& de Jong (2001)                     & 3.53$\pm0.20$&2.73& H{\sevensize I}& H{\sevensize I}+He\\
  &   \vline & Avila-Reese et al. (2008)$^{\textit{a}}$   & 3.15$\pm0.01$&3.59&  H{\sevensize I}&H{\sevensize I}+He+H$_{2}$  \\
  &   \vline & De Rijcke et al.  (2007)                   & 3.15$\pm0.07$&3.25&  H{\sevensize I} &H{\sevensize I}+He\\
  &   \vline & Kassin et al. (2006)$^{\textit{a}}$        & 3.40$\pm0.30$&2.87& H$\alpha$& H{\sevensize I}+He\\
  &   \vline &  Meyer et al. (2008)                       & 4.35$\pm0.14$&2.20& H{\sevensize I}&H{\sevensize I}+He+H$_{2}$ \\
\hline
\multirow{6}{*}{B03}   
  &   \vline & GHASP sample (this work, L$_{K}$)          & 3.64$\pm0.28$&2.21&H$\alpha$ & H{\sevensize I}+He \\
  &   \vline & GHASP sample (this work, L$_{K}$, H$_{2}$) & 3.63$\pm$0.28&2.33&H$\alpha$ & H{\sevensize I}+He+H$_{2}$    \\
  &   \vline & GHASP sample (this work, L$_{B}$, H$_{2}$) & 3.25$\pm0.29$&3.07& H$\alpha$ & H{\sevensize I}+He+H$_{2}$  \\
  &   \vline & McGaugh (2005)                             & 3.19$\pm0.14$&3.23  & H{\sevensize I} & H{\sevensize I}+He\\
  &   \vline & Stark et al. (2009)                        & 3.93$\pm0.07$&1.78   & H{\sevensize I} & H{\sevensize I}+He\\
  &   \vline & Kassin et al. (2006)                       & 3.40$\pm0.30$&2.87    & H{\sevensize I} & H{\sevensize I}+He \\
  &   \vline & Avila-Reese et al. (2008)                  & 3.15$\pm0.01$&3.59&  H{\sevensize I} & H{\sevensize I}+He+H$_{2}$\\
  &   \vline & Geha et al. (2006)                         & 3.70$\pm0.15$&   & H{\sevensize I} & H{\sevensize I}+He\\
\hline
\multirow{2}{*}{P04}       
  &   \vline & GHASP sample (this work, L$_{K}$)          & 3.95$\pm0.31$&1.69&H$\alpha$ & H{\sevensize I}+He \\    
  &   \vline & GHASP sample (this work, L$_{K}$, H$_{2}$) & 3.89$\pm$0.30&1.89&H$\alpha$& H{\sevensize I}+He+H$_{2}$\\
  &   \vline & Stark et al. (2009)                        & 3.94$\pm0.07$&1.79   & H{\sevensize I} & H{\sevensize I}+He\\    
\hline
\multirow{4}{*}{M00$^{\textit{b}}$}   
  &   \vline & GHASP sample (this work, L$_{K}$)          & 3.54$\pm0.37$&2.44  & H$\alpha$ & H{\sevensize I}+He \\ 
  &   \vline & GHASP sample (this work, L$_{K}$, H$_{2}$) & 3.54$\pm$0.37&2.52    &H$\alpha$  & H{\sevensize I}+He+H$_{2}$ \\
  &   \vline & McGaugh et al. (2000)                      & 3.98$\pm0.12$&1.57  & H{\sevensize I} & H{\sevensize I}+He \\ 
  &   \vline & Noordermeer \& Verheijen (2007)            & 3.36$\pm0.10$&    2.14       &  H{\sevensize I} & H{\sevensize I}+He\\
  &   \vline & Pfenniger \& Revaz (2005)                  & 3.36$\pm0.10$&3.11   & H{\sevensize I}& H{\sevensize I}+He\\ 
  &   \vline & Gurovich et al. (2010)                     & 3.20$\pm0.10$&2.50   & H{\sevensize I}& H{\sevensize I}+He\\ 

\hline
\end{tabular}
\\Notes. $^{\textit{a}}$: Stellar mass estimated from B\&J and BE. Coefficients follow $M_{bar}=(c\pm\sigma_{c})V_{max}^{(\alpha\pm\sigma_{\alpha})}$, where log($c\pm\sigma_{c}$)=$\beta\pm\sigma_{\beta}$. $^{\textit{b}}$: In this case we use an unweighted bisector fit (given that we assume a constant $\Upsilon_{\star}$).
\label{tablehenri}
\end{minipage}
\end{table*}

\section{Discussion}

\subsection{The slope of the baryonic Tully-Fisher relation}

In this paper, we have estimated $\Upsilon_{\star}$ for field galaxies
using the usual recipes given in B\&J, BE and PO. We used these values
to compute their stellar and baryonic masses. We found that the slopes
of the stellar and baryonic TF relations do not change significantly,
whatever the method used for computing the masses. We obtain similar results when
we fixed the $\Upsilon_{\star}$ to 0.8 (MG). Table \ref{tablehenri}
summarizes the different slopes of the baryonic TF relation from different
authors using different $\Upsilon_{\star}$ and gas masses. In column 1,
this table lists the method used to estimate the $\Upsilon_{\star}$
(BJ01, B03, P04 and MG). Column 2 presents the different samples of
galaxies and their references with the resulting slopes and zero points
(columns 3 and 4, respectively) using the methods listed in column
1. We computed the results obtained for the GHASP sample for all
$\Upsilon_{\star}$ estimators (and also when the H$_{2}$ was included
in the gaseous mass). We note in columns 5 and 6 which tracer was used
for the rotation curve (derived from H{\sevensize I} or H$\alpha$ data)
and how the gas mass was estimated (H{\sevensize I} masses corrected by helium and metals or H{\sevensize I} masses corrected by helium and metals and molecular hydrogen).

In the next sections, we compare different slopes of the baryonic TF
relation listed in Table \ref{tablehenri}. Except Kassin et al. (2006),
who estimated the maximum rotational velocity from H$\alpha$ data,
all the other studies compute the maximum rotational velocity using
H{\sevensize I} data. In addition to the different models to compute the stellar
mass, the main difference between all studies comes from the sample
used. B\&J used data from cluster galaxies in Ursa Major. Avila-Reese
et al. (2008) based their study on a compilation of samples (Zavala et
al. 2003) including mostly high surface brightness and late-type disk
galaxies complemented with low surface brightness galaxies compiled from
the literature. De Rijcke et al. (2007) used dwarf ellipticals galaxies
mostly located in groups and clusters and complemented this sample with
spirals coming from the study of Tully \& Pierce (2000), McGaugh (2005)
and Geha et al. (2006). Meyer et al. (2008) used the HIPASS catalog plus
2MASS photometry for a large range of morphological types. Noordermeer
et al. (2007) used a sample including massive disk galaxies from the
WHISP survey. McGaugh (2005) used a compilation from Sanders \& McGaugh
(2002), combining late-type galaxies with maximum rotational velocities
ranging from 20 to 300 km s$^{-1}$.

McGaugh (2005) estimated $\Upsilon_{\star}$, in the B-band,  using
a maximum disk model, stellar population synthesis models and the
mass discrepancy-acceleration relation. Considering these stellar
mass-to-light ratios, McGaugh (2005) searched for a slope that minimized
the scatter in the baryonic TF relation by using a $\Upsilon_{\star}$
coming from the mass discrepancy-acceleration relation. These authors
give a slope of $\sim$4. Using the recipe of Bell et al. (2003) to
estimate stellar masses (with a Kroupa IMF), Geha et al. (2006) found a
slope of 3.70$\pm$0.15. Using a fixed mass-to-light ratio for a specific
band, McGaugh (2005) found a slope of $\sim$4 for the baryonic TF relation using
galaxies having rotational velocities between 30$\leq$V$_{rot}\leq$300
km s$^{-1}$. Stark et al. (2009) and Trachternach et al. (2009) used
stellar population synthesis models and obtained a slope of $\sim$4.

On the other hand, De Rijcke et al. (2007) found a slope of 3.15$\pm$0.07
for the baryonic TF relation for a sample of early and late-type
galaxies. McGaugh (2005) found a similar slope of $\alpha=3.19\pm0.14$
when he derived stellar masses using stellar population synthesis
models of Bell et al. (2003). Fixing the $\Upsilon_{\star}$ to 0.8,
Noordermeer \& Verheijen (2007) found slopes ranging from 3.04$\pm$0.08
to 3.38$\pm$0.10, depending of the adopted rotational velocity. Kassin
et al. (2006) found slopes of 3.3$\pm$0.3 and 3.1$\pm$0.3, also depending
on the adopted rotational velocity.

Using the method given in BE for the stellar mass determination,
the GHASP slope of the baryonic Tully-Fisher relation is
$\alpha=3.52\pm0.27$. Taking into account the uncertainties, we found that this
value is in agreement with most of the results listed above.

As already mentioned  in \S 2.3, there is no consensus in the literature about the correlation 
that should exist between dynamical and stellar populations
synthesis estimations of $\Upsilon_{\star}$. On one hand, Barnes et
al. (2004) did not find a clear trend between the $\Upsilon_{\star}$ and
(B-R) color (as predicted by Bell et al. (2003), while de Block et al. (2008) and Williams et al. (2009),
found one. These differences could depend on the dynamical mass model used. Barnes et al. (2004) argue that it is difficult to have a reliable estimate of $\Upsilon_{\star}$ without a good knowledge of the dark matter content of galaxies. On the
other hand, Williams et al. (2009) do not use rotation curves in order
to derive $\Upsilon_{\star}$, but a dynamical stellar modeling
and compare it with observed rotation curves. Williams et al. (2009) found
a median $\Upsilon_{\star}$ of 1.09 in the K-band. If we use the same
value to obtain the slope of the baryonic TF relation for the GHASP
sample, we found a slope of 3.61$\pm$0.37 in good agreement with all
the estimations of the slope using stellar population models.

\subsection{Baryonic, stellar and total masses in the GHASP sample}

The slope of the baryonic TF relation is a test for galaxy formation
models, given that they try to reproduce the main properties of observed galaxies
(e. g. Portinari et al. 2007). Fig. \ref{totmassbary} illustrates that
a steeper slope in the baryonic TF relation indicates that high mass
galaxies tend to have their baryonic masses better matching their total masses while, in contrast, the fraction of baryonic mass over the total
mass decreases for low-mass galaxies. In the case of GHASP galaxies,
low-mass galaxies (V$_{max}<$150 km s$^{-1}$) seem to be dominated
by dark matter (corresponding to 71$\%$ of their total masses), while intermediate and
high-mass spirals (V$_{max}>$150 km s$^{-1}$) are baryonic matter
dominated (corresponding to 37$\%$ of their total masses).

Due to the simply model we used to compute the total dynamical masses, these baryonic mass fractions are probably lower limits. Indeed, the total mass within the optical radius is distributed between a spheroidal dark matter halo and a thin optical disk and not simply within a simple spherical distribution.  For a given galaxy mass enclosed in a spheroid, the circular speed decreases by $\sim20\%$ when the axis ratio of the spheroid increases from 0.1 (disk case) to 1.0 (spherical case, Binney \& Tremaine 2008, Lequeux 1983). Thus, to take into account the geometry of the system composed of a flat disk and a spherical halo, we have estimated the total dynamical mass should be lowered by a factor depending on the disk-to-halo mass ratio. This increases the fraction of baryonic matter to 31\% for low mass galaxies and to 38\% for intermediate and high mass systems. Both approaches lead to the same main conclusion: low-mass galaxies have lower baryonic fractions than high-mass galaxies.

The question whether dark matter is coupled with some baryonic mass component of galaxies is
still unanswered. Pfenniger, Combes \& Martinet (1994) suggest that cold molecular gas could explain the dark matter content in galaxies. Hoekstra et al. (2001) have used a scaling in the neutral hydrogen of galaxies to explain the dark matter in spiral galaxies. In the optical, this analysis is difficult to lead due to the limited extension of optical rotation curves versus H{\sevensize I} ones.

In Figs. \ref{r25bary}, \ref{totmassbary} and \ref{BTF_GHASP_fig5} we
have shown relations between the total and baryonic masses and how these
values are related with radius and $\Upsilon_{\star}$. As expected,
we found that larger galaxies exhibit larger amounts of baryonic matter
(e. g. Kassin et al. 2006), suggesting that the baryonic mass density
($\Sigma_{bar}$) is weakly dependent on the size of the galaxy.

An important issue related to the Tully-Fisher relation is the photometric
band used in its construction. B\&J found that the stellar and/or
baryonic Tully-Fisher relations are independent of the passband used. This
result is expected and very important, because we have to consider the
total mass of a galaxy in the baryonic mass budget. In our case,
we found a difference in the slope of the baryonic Tully-Fisher relation
when the stellar mass is calculated from B or K-band luminosities (see
Table \ref{tablehenri}). There are a number of possible causes. One possible explanation could come from the way that luminosities are computed. In the case of the K-band, we have used
the total magnitude at the isophote of 20 mag arcsec$^{-2}$, while in
the case of the B-band, we have used \textit{SDSS} g-band magnitudes, which were converted into B-band. To check this issue, we also used a Kron
elliptical aperture magnitude to estimate the K-band luminosities. Using
this alternative method, the slope of the baryonic Tully-Fisher relation
is in agreement with the value obtained by using the luminosities at the
isophote 20 mag arcsec$^{2}$ (the difference between both estimations
is about $\sim$0.05) and therefore the difference with the
B-band analysis remains.

Additional reasons why the B and K-band slopes do not agree may be: (1) the extinction corrections between both bands are quite different and may be incorrect; (2) the $\Upsilon_{\star}$ prescriptions of the various groups which differ substantially, especially in the K-band and other near IR passbands (for instance assumptions for star formation histories and metallicity spread may differ) and (3) the stellar  population synthesis models that underlie the SED-$\Upsilon_{\star}$ prescriptions  show substantial differences in the near-IR, because these passbands are dominated by late stages of stellar evolution that are still poorly understood. Most notably is  the importance of TP-AGB stars, discussed in \S3.1, but also other evolved stages could play a role.

\subsection{Cosmological predictions of the baryonic Tully-Fisher relation}

In the past years, several studies have been devoted to study the Tully-Fisher
relation using cosmological simulations (e. .g Bullock et al. 2001,
Governato et al. 2007, Portinari \& Sommer-Larsen 2007 and Piontek \&
Steinmetz 2009). These authors found slopes shallower than 4 for the
baryonic TF relation. For instance, Bullock et al. (2001) found a slope
of 3.40$\pm$0.05 and Steinmetz \& Navarro (1999) suggest a slope of 3
in the standard CDM universe. Portinari \& Sommer-Larsen (2007) found
an excellent agreement between the predicted baryonic TF relation and
the relation obtained by McGaugh (2005) using stellar populations. Our
slope $\alpha=3.64\pm0.28$ is in agreement (in one-$\sigma$) with the results shown by
Bullock et al. (2001). We note that the slope of the baryonic TF for the
GHASP sample ($\alpha=3.64\pm0.28$) is steeper than the value expected from the virial theorem:
$M_{vir} \propto V^{3}_{vir}$, however, a steeper slope is expected when
the concentration of the halo is taken into account (Bullock et al. 2001).

Several authors have suggested that low-mass galaxies could have higher
ratios of dark-to-luminous ratios (Tinsley 1981, Persic, Salucci \&
Stel 1996, C\^ot\'e, Carignan, \& Freeman 2000). Some of the physical
processes associated with that phenomena could be supernova feedback
(van den Bosch 2000) and ram pressure stripping of gas (Mori \& Burkert
2000). Using morphological types, we have shown that late-type spirals in the GHASP sample have higher total-to-baryonic masses ratio than early-type spirals (even when H$_{2}$ gas mass is included in the baryonic mass budget, see Fig. \ref{BTF_GHASP_fig5}). This is in quantitative agreement with most current galaxy formation models under the assumption that the $\Upsilon_{\star}$ prescriptions are applicable and correct over the full mass range of galaxies in the GHASP sample.

\section{Summary and conclusions}

The main focus of this paper is the study of the NIR, stellar and baryonic Tully-Fisher relation for an homogeneous sample of local galaxies, and determine their implications in the total mass of galaxies. In that frame, we have used the survey GHASP. Given that maximum velocities for GHASP galaxies were derived from rotation curves computed from 2D velocity fields, we were able to minimize the uncertainties. This allows us to determine the maximum rotational velocity of galaxies without any previous assumption about the inclination and position angle of the object. Thus, the GHASP survey is an ideal laboratory to study the Tully-Fisher relation in an homogenous way.

In the following, we list the main findings of this work:

\begin{enumerate}

\item We estimated the slope of the NIR, stellar and baryonic Tully-Fisher relation for the GHASP survey. We have found a slope of $\alpha=3.64\pm0.28$ for the baryonic Tully-Fisher relation, as estimated by using the BE model. In the case of the baryonic relation, we found that whatever model is used to estimate $\Upsilon_{\star}$ (B\&J, BE, PO or MG), the slope of the baryonic Tully-Fisher relation does not change significantly. Recent cold dark matter galaxy formation models are consistent with our results.

\item The baryonic mass of a galaxy grows almost linearly with its optical radius. This means that the surface baryonic mass density is weakly dependent on the size of the galaxies ($\Sigma_{b} \propto R_{25}^{0.4}$).

\item We have classified the shape of the rotation curves in order to study their influence on the Tully-Fisher relation. We found that rising rotation curves tend to be farther from the fit than flat rotation curves. We found that galaxies having asymmetric rotation curves, or alternatively galaxies showing larger residual velocities, are also the galaxies presenting the higher dispersion in the Tully-Fisher relation. The fact that only the dispersion of the Tully-Fisher relation (and not the slope or the zeropoint) is affected by the shape and the asymmetries of the rotation curves shows that maximum rotation velocity determinations are rather robust regardless the quality of the rotation curve.

\end{enumerate}

Here we list the results that we have strengthened and clarified by using the GHASP survey:

\begin{enumerate}

\item We bear out the presence of a break in the NIR Tully-Fisher relation at M$_{H,K}\sim$-20 in the sense of low-luminosity galaxies being less luminous (or having higher rotational velocities) than the values expected from the Tully-Fisher relation defined by high-mass galaxies.

\item Taken into account the uncertainties, the slope of the baryonic Tully-Fisher relation for the GHASP survey is in agreement with the slope found by Bell \& de Jong (2001), Kassin et al. (2006), Geha et al. (2006).

\item This work supports that late-type spiral galaxies (which are usually
low-mass galaxies) present higher total-to-baryonic mass ratios than early-type spiral galaxies, in
agreement with previous observations and with cold dark matter simulations. In
this sense, high (low) total-to-baryonic mass ratio may be explained
either by a high (low) dark matter content or (and) low (high) baryonic
content. Low (high) baryonic content could be explained by the escape
of a fraction of the baryons to the intergalactic medium (internal feedback). Alternatively there could be the case that the baryons never collapsed that much into the centers of low mass dark matter halos, i.e., that high mass galaxies have more concentrated baryon distributions than low mass galaxies for this reason.

\end{enumerate}
 
\section*{acknowledgments}

We would like to thank the anonymous referee for the very useful comments that improved this paper considerably. This publication makes use of data products from the Two Micron All
Sky Survey, which is a joint project of the University of Massachusetts
and the Infrared Processing and Analysis Center/California Institute of
Technology, funded by the National Aeronautics and Space Administration
and the National Science Foundation. S. T--F. acknowledges the financial support of FONDECYT through a post-doctoral position, under contract 3110087 and also to FAPESP (doctoral fellowship, under
contract 2007/07973-3). S. T-F also acknowledges the financial support
of EGIDE through an Eiffel scholarship and also the ARCUS program. CMdO
acknowledges support from FAPESP (2006/56213-9) and CNPq. HP acknowledges
financial support from CNPq (201600/2009-9 and 471254/2008-8). 
We also acknowledge the usage of the HyperLeda database (http://leda.univ-lyon1.fr). S. T--F. and HP would like to thank the staff members of the Laboratoire d'Astrophysique de Marseille for their hospitality when part of this work was developed.

\appendix

\section{Tables} 

\begin{table*}
\begin{minipage}[t]{\textwidth}
\caption{NIR photometry and rotational velocities of the sample}
\label{appendix1}
\centering
\renewcommand{\footnoterule}{}  % to avoid a line before footnotes
\begin{tabular}{ccccccc}
%\hline
%& & & &\multicolumn{4}{c}{Observed} &\multicolumn{1}{c}{Model} \\
\hline
Galaxy & M${\rm_{H}}$ & M${\rm_{K}}$ & V$_{\rm{max}}^{\rm{RC}}$ & V$_{\rm{R(25)}}^{\rm{model}}$ & V$_{\rm{max}}^{\rm{TF}}$ &Flags \\  
UGC&mag&mag& km s$^{-1}$ & km s$^{-1}$ & km s$^{-1}$ & \\
(1)&(2)&(3)&(4)&(5)&(6)&(7)\\ 
\hline  	    
89     &-24.90 & -25.20  & 343$\pm$117& 340 &	 343$\pm$117 &  F--\\
94     &-23.16 & -23.40  & 209$\pm$21 & 210 &	 209$\pm$21  &  F \\
763    &-20.54 & -20.74  & 104$\pm$11 & 95  &	 104$\pm$11  &  F+\\
1256   &-20.30 & -20.43  & 105$\pm$9  & 119 &	 105$\pm$9   &  R+\\
1317   &-24.44 & -24.69  & 205$\pm$9  & 208 &	 205$\pm$9   &  F\\
1437   &-24.42 & -24.73  & 218$\pm$15 & 210 &	 218$\pm$15  &  D+\\
1886   &-24.29 & -24.53  & 267$\pm$8  & 257 &	 267$\pm$8   &  F+ \\
2141   &-20.98 & -21.11  & 157$\pm$20 & 120 &	 157$\pm$20  &  R+\\
2183   &-22.00 & -22.26  & 160$\pm$32 & 164 &	 164$\pm$25* &  F-- \\
2503   &-24.43 & -24.66  & 285$\pm$12 & 291 &	 285$\pm$12  &  F\\
2800   &-20.25 & -20.45  & 103$\pm$20 & 108 &	 108$\pm$13* &  R-- \\
2855   &-24.06 & -24.24  & 229$\pm$9  & 221 &	 229$\pm$9   &  F+ \\
3273   &-19.07 & -19.25  & 106$\pm$7  & 81  &	 106$\pm$7   &  R+\\
3334   &-25.61 & -25.89  & 377$\pm$85 & 385 &	 385$\pm$62* &  F--\\
3429   &-23.89 & -24.23  & 322$\pm$30 & 274 &	 322$\pm$30  &  R--\\
3521   &-22.48 & -22.74  & 166$\pm$12 & 169 &	 169$\pm$8*  &  F--\\
3528   &-23.53 & -23.79  & 276$\pm$66 & 273 &	 276$\pm$66  &  F-- \\
3685   &-22.17 & -22.40  & 133$\pm$177& 105 &	 133$\pm$177 &  R+\\
3691   &-21.47 & -21.52  & 143$\pm$10 & 131 &	 143$\pm$10  &  R+ \\
3709   &-24.33 & -24.65  & 241$\pm$14 & 240 &	 241$\pm$14  &  F\\
3734   &-21.36 & -21.59  & 108$\pm$16 & 94  &	 108$\pm$16  &  R+ \\
3740   &-21.64 & -21.97  & 87 $\pm$20 & 83  &	  87$\pm$20  &  R+\\
3915   &-23.31 & -23.60  & 205$\pm$16 & 212 &	 212$\pm$12* &  F-- \\
4026   &-24.10 & -24.34  & 284$\pm$14 & 282 &	 284$\pm$14  &  F-- \\
4273   &-22.98 & -23.22  & 219$\pm$11 & 191 &	 219$\pm$11  &  R+ \\
4284   &-19.19 & -19.33  & 118$\pm$14 & 103 &	 118$\pm$14  &  R+\\
4325   &-16.51 & -16.70  &  85$\pm$13 & 88  &	  88$\pm$7*  &  R--\\
4820   &-23.76 & -24.10  & 336$\pm$20 & 339 &	 339$\pm$21* &  F--\\
5175   &-23.24 & -23.43  & 188$\pm$10 & 199 &	 199$\pm$6*  &  R-- \\
5251   &-21.74 & -21.92  & 125$\pm$9  & 123 &	 125$\pm$9   &  F--\\
5316   &-19.88 & -20.05  & 145$\pm$9  & 108 &	 145$\pm$9   &  R+\\
5351   &-22.35 & -22.58  & 135$\pm$8  & ... &	 135$\pm$8   & ... \\
5721   &-17.87 & -18.02  & 99 $\pm$29 & 74  &	  99$\pm$29  &  R+\\
5789   &-19.82 & -20.03  & 131$\pm$10 & 92  &	 131$\pm$10  &  R+\\
5842   &-21.12 & -21.35  & 115$\pm$18 & 108 &	 115$\pm$18  &  R \\
5931   &-21.46 & -21.65  & 157$\pm$32 & 126 &	 157$\pm$32  &  R+\\
5982   &-22.45 & -22.66  & 199$\pm$13 & 184 &	 199$\pm$13  &  F--\\
6118   &-22.93 & -23.22  & 137$\pm$24 & 133 &	 137$\pm$24  &  R--\\
6521   &-23.95 & -24.18  & 249$\pm$18 & 252 &	 252$\pm$15* &  F--\\
6537   &-22.63 & -22.79  & 187$\pm$17 & 197 &	 187$\pm$17  &  D \\
6628   &-17.86 & -18.09  & 183$\pm$168& 151 &	 183$\pm$168 &  R+\\
6702   &-23.54 & -23.91  & 195$\pm$23 & 187 &	 195$\pm$23  &  F+\\
6778   &-22.80 & -23.00	 & 223$\pm$14 & 187 &	 223$\pm$14  &  R+ \\
7045   &-21.98 & -22.19  & 160$\pm$9  & 156 &	 160$\pm$9   &  R+ \\
7831   &-20.61 & -20.77  &  92$\pm$15 & 98  &	  98$\pm$10* &  R-- \\
7861   &-20.21 & -20.44  &  50$\pm$21 & 33  &	  50$\pm$21  &  R \\
7901   &-23.31 & -23.53  & 215$\pm$10 & 218 &	 215$\pm$10  &  F \\
8403   &-20.90 & -21.07  & 128$\pm$10 & 118 &	 128$\pm$10  &  R+ \\
8490   &-18.09 & -18.28  &  90$\pm$29 & 89  &	  90$\pm$29  &  R--\\
8852   &-23.06 & -23.28  & 186$\pm$10 & 195 &	 195$\pm$7*  &  F--\\
8900   &-24.60 & -24.87  & 345$\pm$37 & 361 &	 361$\pm$31* &  R-- \\
9179   &-18.60 & -18.70  & 111$\pm$36 & 105 &	 111$\pm$36  & F+ \\
9248   &-22.80 & -23.05  & 166$\pm$11 & 179 &	 166$\pm$11  &  F \\
9366   &-24.74 & -24.98  & 241$\pm$9  & 238 &	 241$\pm$9   &  F--\\
9576   &-21.00 & -21.15  & 104$\pm$25 & 91  &	 104$\pm$25  &  F+\\
9736   &-22.64 & -22.90  & 192$\pm$16 & 187 &	 192$\pm$16  & R--\\
9753   &-21.50 & -21.69  & 138$\pm$9  & 146 &	 146$\pm$1*  &  F-- \\
9866   &-19.84 & -20.02  & 114$\pm$11 & 115 &	 115$\pm$7*  &  R-- \\
\hline
\end{tabular}
\end{minipage}
\end{table*}

\begin{table*}
\begin{minipage}[t]{\textwidth}
\caption{...continued}
\label{table4}
\centering
\renewcommand{\footnoterule}{}  % to avoid a line before footnotes
\begin{tabular}{ccccccc}
%\hline
%& & & &\multicolumn{4}{c}{Observed} &\multicolumn{1}{c}{Model} \\
\hline
Galaxy & M${\rm_{H}}$ & M${\rm_{K}}$  & V$_{\rm{max}}^{\rm{RC}}$ & V$_{\rm{R(25)}}^{\rm{model}}$ & V$_{\rm{max}}^{\rm{TF}}$ &Flags \\  
UGC& mag&mag& km s$^{-1}$ & km s$^{-1}$ & km s$^{-1}$ &  \\
(1)\footnotetext{Column (1): Galaxy name. Column (2): H-band absolute magnitude. Column (3): K-band absolute magnitude. Column (4): Observed maximum rotational velocity. Column (5): Rotational velocity at R$_{25}$ derived from the arctan model. In the case of UGC 5351, the maximum velocity was obtained from the position velocity diagram, therefore, no arctan model could be fitted for this object. Column (6): Rotational velocities used in the TF relation. An asterisk marks the rotational velocities derived from the arctan model. In these cases, the uncertainties were computed from the kinematical inclination given in Epinat et al. (2008b). Column (7): Flag on the rotation curve (as described in the \S2.1.)}&(2)&(3)&(4)&(5)&(6)&(7)\\ 
\hline  	    
9969  &-24.32 & -24.52 &   311$\pm$9  & 321 &  311$\pm$9  &  2F\\
10075 &-22.03 & -22.25 &   168$\pm$9  & 170 &  168$\pm$9  &  1F-- \\
10445 &-17.97 & -18.08 &   77 $\pm$17 & 83  &	77$\pm$17 &  1D+ \\
10470 &-22.69 & -22.92 &   164$\pm$39 & 160 &  164$\pm$39 &  1F+\\
10502 &-23.21 & -23.42 &   163$\pm$14 & 157 &  163$\pm$14 &  2F+ \\
10521 &-21.88 & -22.08 &   124$\pm$9  & 128 &  128$\pm$4*  & 1R-- \\
10546 &-19.55 & -19.72 &   106$\pm$22 & 106 &  106$\pm$22 &  2F+ \\
10564 &-18.10 & -18.17 &   75 $\pm$8  & 51  &	75$\pm$8  &  2R+ \\
10757 &-18.78 & -19.07 &   81 $\pm$33 & 80  &	81$\pm$33 &  2R\\
10897 &-21.62 & -21.84 &   113$\pm$56 & 117 &  117$\pm$35* & 1R--\\
11012 &-21.20 & -21.37 &   117$\pm$9  & 127 &  127$\pm$1* &  1F--\\
11218 &-23.37 & -23.60 &   185$\pm$9  & 188 &  188$\pm$4*  & 1F-- \\
11269 &-23.10 & -23.41 &   202$\pm$13 & 197 &  202$\pm$13 &  3F+ \\
11283 &-19.50 & -19.52 &   173$\pm$73 & 160 &  173$\pm$73 &  2R\\
11300 &-19.32 & -19.41 &   112$\pm$9  & 93  &  112$\pm$9  &  1R\\
11429 &-24.73 & -25.00 &   232$\pm$35 & 208 &  232$\pm$35 &  3R+ \\
11498 &-24.36 & -24.62 &   274$\pm$9  & 280 &  274$\pm$9  &  3F+ \\
11852 &-23.56 & -23.86 &   221$\pm$27 & 247 &  247$\pm$21* & 3R-- \\
11861 &-22.36 & -22.51 &   181$\pm$39 & 134 &  181$\pm$39 &  1R+ \\
11872 &-22.91 & -23.13 &   183$\pm$12 & 188 &  188$\pm$8* &  1F--\\
11914 &-23.80 & -24.03 &   285$\pm$26 & 283 &  283$\pm$27* & 2D-- \\
11951 &-21.09 & -21.29 &   106$\pm$7  & 102 &  106$\pm$7  &  1R+ \\
12276 &-23.55 & -23.78 &   94 $\pm$37 & 93  &	94$\pm$37 &  2F+\\
12343 &-23.74 & -23.96 &   221$\pm$14 & 243 &  221$\pm$14 &  2F+\\
12754 &-19.64 & -19.89 &   123$\pm$11 & 122 &  123$\pm$11 &  1F+\\
\hline
\end{tabular}
\end{minipage}
\end{table*}

\begin{table*}
\centering
\begin{minipage}[t]{\textwidth}
\caption{Main parameters of the GHASP sample}
\centering
\begin{tabular}{ccccccccccc}
\hline
\multicolumn{1}{c}{Galaxy}& Radius&B-V & $\Upsilon_{K}^{B\&J}$ & $\Upsilon_{K}^{BE}$ &$\Upsilon_{K}^{PO}$ & log M$_{\star}^{B\&J}$& log M$_{\star}^{BE}$ & log M$_{\star}^{PO}$ & log M$_{gas}$& log M$_{bar}$\\
\multicolumn{1}{c}{UGC} & Kpc &mag & & & &M$_\odot$ & M$_\odot$ &M$_\odot$ & M$_\odot$ & M$_\odot$ \\
\multicolumn{1}{c}{(1)}&(2)&(3)&(4)&(5)&(6)&(7)&(8) &(9) &(10)&(11)\footnotetext{Column (1): Galaxy identification. Column (2): Optical radius taken from \textit{SDSS}. This radius corresponds to the semimajor axis of the isophote of 25 mag arcsec$^{-2}$ (See \S 2.2). Column (3): Colors B-V (transformed from \textit{SDSS} g-r colors). Columns (4), (5) and (6) list the mass-to-light ratios calculated from B\&J, BE and PO, respectively. In columns (7), (8) and (9) we list the stellar masses derived from B\&J, BE and PO, respectively. Column (10) lists the H{\sevensize I} masses, corrected by helium and metals. Column (11) correspond to the baryonic mass (M$_{\star}$+M$_{gas}$) where M$_{\star}$ was calculated using BE.}\\
 \hline
   89&    23.30   & 0.85 &0.73 & 0.81& 1.18 & 11.31$\pm$0.10 & 11.35$\pm$0.10 &11.52$\pm$0.10  &  9.92$\pm$0.05  & 11.37$\pm$0.10    \\
   94& ...        & 0.78 &0.66 & 0.79& 1.04 & 10.54$\pm$0.10 & 10.62$\pm$0.10 &10.74$\pm$0.10  & 10.03$\pm$0.05  & 10.72$\pm$0.08    \\
  763& ...        & 0.77 &0.65 & 0.79& 1.03 &  9.47$\pm$0.10 &  9.56$\pm$0.10 & 9.67$\pm$0.10  &  9.45$\pm$0.04  &  9.81$\pm$0.06    \\
 1317&    26.80   & 1.00 &0.91 & 0.85& 1.50 & 11.20$\pm$0.10 & 11.17$\pm$0.10 &11.42$\pm$0.10  & 10.35$\pm$0.04  & 11.23$\pm$0.09    \\
 2141&     4.65   & 0.68 &0.56 & 0.77& 0.87 &  9.56$\pm$0.10 &  9.69$\pm$0.10 & 9.75$\pm$0.10  &  9.24$\pm$0.04  &  9.82$\pm$0.08  \\
 2183&     5.89   & 0.99 &0.90 & 0.85& 1.47 & 10.22$\pm$0.10 & 10.20$\pm$0.10 &10.44$\pm$0.10  &  9.56$\pm$0.10  & 10.29$\pm$0.08    \\
 3521&    15.65   & 0.77 &0.64 & 0.79& 1.01 & 10.27$\pm$0.11 & 10.36$\pm$0.11 &10.47$\pm$0.11  &  9.97$\pm$0.05  & 10.51$\pm$0.08    \\
 4026&    21.78   & 0.97 &0.87 & 0.84& 1.43 & 11.04$\pm$0.10 & 11.03$\pm$0.10 &11.26$\pm$0.10  &  9.49$\pm$0.17  & 11.04$\pm$0.10    \\
 4273&    15.99   & 0.95 &0.85 & 0.84& 1.39 & 10.58$\pm$0.10 & 10.57$\pm$0.10 &10.79$\pm$0.10  &  9.90$\pm$0.14  & 10.66$\pm$0.09    \\
 4284&     4.87   & 0.77 &0.65 & 0.79& 1.03 &  8.91$\pm$0.10 &  8.99$\pm$0.10 & 9.11$\pm$0.10  &  9.46$\pm$0.03  &  9.59$\pm$0.03    \\
 4325&     3.80   & 0.87 &0.75 & 0.82& 1.21 &  7.92$\pm$0.12 &  7.96$\pm$0.12 & 8.13$\pm$0.12  &  8.91$\pm$0.03  &  8.96$\pm$0.03    \\
 4820&    10.64   & 0.92 &0.81 & 0.83& 1.31 & 10.91$\pm$0.10 & 10.92$\pm$0.10 &11.12$\pm$0.10  &  8.65$\pm$0.09  & 10.92$\pm$0.10    \\
 5351&     8.79   & 0.92 &0.81 & 0.83& 1.31 & 10.30$\pm$0.10 & 10.31$\pm$0.10 &10.51$\pm$0.10  &  8.80$\pm$0.06  & 10.33$\pm$0.10    \\
 5721& ...        & 0.63 &0.52 & 0.76& 0.81 &  8.29$\pm$0.10 &  8.45$\pm$0.10 & 8.48$\pm$0.10  &  8.73$\pm$0.06  &  8.91$\pm$0.05     \\
 5789& ...        & 0.64 &0.53 & 0.76& 0.82 &  9.10$\pm$0.10 &  9.26$\pm$0.10 & 9.29$\pm$0.10  &  9.56$\pm$0.03  &  9.74$\pm$0.04    \\
 5842&     8.30   & 0.88 &0.76 & 0.82& 1.23 &  9.79$\pm$0.10 &  9.82$\pm$0.10 & 9.99$\pm$0.10  &  9.03$\pm$0.05  &  9.88$\pm$0.09    \\
 5982&    11.75   & 0.93 &0.82 & 0.83& 1.33 & 10.34$\pm$0.10 & 10.35$\pm$0.10 &10.55$\pm$0.10  &  9.72$\pm$0.06  & 10.44$\pm$0.08     \\
 6118&     7.33   & 0.79 &0.67 & 0.80& 1.06 & 10.48$\pm$0.10 & 10.55$\pm$0.10 &10.68$\pm$0.10  &  8.79$\pm$0.08  & 10.56$\pm$0.10    \\
 6521&    27.69   & 0.84 &0.71 & 0.81& 1.14 & 10.89$\pm$0.10 & 10.94$\pm$0.10 &11.09$\pm$0.10  & 10.15$\pm$0.04  & 11.01$\pm$0.09    \\
 6628&     4.42   & 0.69 &0.58 & 0.77& 0.90 &  8.36$\pm$0.11 &  8.49$\pm$0.11 & 8.55$\pm$0.11  &  9.14$\pm$0.04  &  9.23$\pm$0.04    \\
 6702&    21.07   & 0.81 &0.69 & 0.80& 1.09 & 10.76$\pm$0.10 & 10.83$\pm$0.10 &10.97$\pm$0.10  &  9.91$\pm$0.06  & 10.88$\pm$0.09     \\
 6778&     8.09   & 0.84 &0.71 & 0.81& 1.14 & 10.42$\pm$0.10 & 10.47$\pm$0.10 &10.62$\pm$0.10  &  9.62$\pm$0.04  & 10.53$\pm$0.09    \\
 7045&     3.48   & 1.66 &2.45 & 1.04& 4.54 & 10.63$\pm$0.10 & 10.26$\pm$0.10 &10.90$\pm$0.10  &  8.93$\pm$0.04  & 10.28$\pm$0.10    \\
 7831&     1.26   & 0.66 &0.55 & 0.76& 0.85 &  9.41$\pm$0.10 &  9.55$\pm$0.10 & 9.60$\pm$0.10  &  8.53$\pm$0.04  &  9.59$\pm$0.09    \\
 7861&     2.90   & 0.71 &0.59 & 0.78& 0.93 &  9.31$\pm$0.10 &  9.43$\pm$0.10 & 9.51$\pm$0.10  &  8.86$\pm$0.05  &  9.53$\pm$0.08    \\
 7901&    12.77   & 0.86 &0.74 & 0.81& 1.20 & 10.65$\pm$0.10 & 10.69$\pm$0.10 &10.85$\pm$0.10  &  9.82$\pm$0.04  & 10.74$\pm$0.09    \\
 8403& ...        & 0.82 &0.70 & 0.80& 1.11 &  9.63$\pm$0.10 &  9.70$\pm$0.10 & 9.84$\pm$0.10  &  9.61$\pm$0.04  &  9.95$\pm$0.06    \\
 8852&     7.83   & 0.94 &0.84 & 0.83& 1.37 & 10.60$\pm$0.10 & 10.60$\pm$0.10 &10.81$\pm$0.10  &  8.68$\pm$0.04  & 10.60$\pm$0.10     \\
 8900&    31.39   & 0.99 &0.90 & 0.85& 1.47 & 11.26$\pm$0.10 & 11.24$\pm$0.10 &11.48$\pm$0.10  & 10.33$\pm$0.08  & 11.29$\pm$0.09    \\
 9179&     3.17   & 0.63 &0.52 & 0.76& 0.81 &  8.56$\pm$0.10 &  8.72$\pm$0.10 & 8.75$\pm$0.10  &  8.96$\pm$0.04  &  9.16$\pm$0.05    \\
 9248&    16.46   & 0.78 &0.66 & 0.79& 1.04 & 10.40$\pm$0.10 & 10.48$\pm$0.10 &10.60$\pm$0.10  &  9.68$\pm$0.04  & 10.55$\pm$0.09    \\
 9366&    29.88   & 0.97 &0.87 & 0.84& 1.43 & 11.30$\pm$0.10 & 11.28$\pm$0.10 &11.51$\pm$0.10  & 10.08$\pm$0.04  & 11.31$\pm$0.10    \\
 9576&    12.05   & 0.67 &0.55 & 0.77& 0.86 &  9.57$\pm$0.11 &  9.71$\pm$0.11 & 9.76$\pm$0.11  &  9.83$\pm$0.06  & 10.07$\pm$0.06    \\
 9736&    11.08   & 0.89 &0.77 & 0.82& 1.25 & 10.41$\pm$0.10 & 10.44$\pm$0.10 &10.62$\pm$0.10  &  9.67$\pm$0.07  & 10.51$\pm$0.09    \\
 9753&     9.06   & 0.87 &0.75 & 0.82& 1.21 &  9.92$\pm$0.10 &  9.95$\pm$0.10 &10.12$\pm$0.10  &  9.09$\pm$0.04  & 10.01$\pm$0.09    \\
 9866&     3.57   & 0.80 &0.68 & 0.80& 1.08 &  9.20$\pm$0.10 &  9.27$\pm$0.10 & 9.40$\pm$0.10  &  7.93$\pm$0.04  &  9.29$\pm$0.10    \\
 9969& ...        & 0.97 &0.87 & 0.84& 1.43 & 11.11$\pm$0.10 & 11.10$\pm$0.10 &11.33$\pm$0.10  & 10.04$\pm$0.06  & 11.13$\pm$0.09    \\
10075&    10.23   & 0.89 &0.77 & 0.82& 1.25 & 10.15$\pm$0.10 & 10.18$\pm$0.10 &10.36$\pm$0.10  &  9.63$\pm$0.03  & 10.29$\pm$0.08    \\
10445&     4.31   & 0.68 &0.56 & 0.77& 0.87 &  8.35$\pm$0.11 &  8.48$\pm$0.11 & 8.54$\pm$0.11  &  9.28$\pm$0.04  &  9.35$\pm$0.03    \\
10757&     4.59   & 0.61 &0.51 & 0.75& 0.79 &  8.70$\pm$0.11 &  8.87$\pm$0.11 & 8.89$\pm$0.11  &  9.11$\pm$0.04  &  9.31$\pm$0.05    \\

\hline
\label{appendix2}
\end{tabular}
\end{minipage}
\end{table*}

\label{lastpage}
\end{document}